\documentclass[aps,prl,superscriptaddress,letterpaper,showpacs,amsmath,amssymb,reprint,floatfix,longbibliography]{revtex4-2}%
\usepackage{graphicx}
\usepackage{stmaryrd}
\usepackage{amsfonts}
\usepackage{siunitx}
\usepackage{physics} % to use \dd
\usepackage[colorlinks=true,citecolor=blue,urlcolor=red]{hyperref}
\usepackage{orcidlink}
\usepackage{soul}

\begin{document}
\title{Wavefront shaping enhanced nano-optomechanics down to the quantum precision limit}

\author{A. G. Tavernarakis\,\orcidlink{0000-0002-9780-2514}}
\affiliation{Université Paris-Saclay, CNRS, ENS Paris-Saclay, CentraleSupélec, LuMIn, 91405, Orsay,
France}

\author{R. Gutiérrez-Cuevas\,\orcidlink{0000-0002-3451-6684}}
\affiliation{Institut Langevin, ESPCI Paris, Université PSL, CNRS, 75005, Paris, France}
\affiliation{Université Paris-Saclay, CNRS, ENS Paris-Saclay, CentraleSupélec, LuMIn, 91405, Orsay,
France}

\author{L. Rondin\,\orcidlink{0000-0002-4833-2886}}
\affiliation{Université Paris-Saclay, CNRS, ENS Paris-Saclay, CentraleSupélec, LuMIn, 91405, Orsay,
France}

\author{T. Antoni\,\orcidlink{0000-0002-0697-0986}}
\affiliation{Université Paris-Saclay, CNRS, ENS Paris-Saclay, CentraleSupélec, LuMIn, 91405, Orsay,
France}

\author{S. M. Popoff\,\orcidlink{0000-0002-7199-9814}}
\affiliation{Institut Langevin, ESPCI Paris, Université PSL, CNRS, 75005, Paris, France}

\author{P. Verlot\,\orcidlink{0000-0002-5105-3319}}
\email{pierre.verlot@universite-paris-saclay.fr}
\affiliation{Université Paris-Saclay, CNRS, ENS Paris-Saclay, CentraleSupélec, LuMIn, 91405, Orsay,
France}
\affiliation{Institut Universitaire de France, 1 rue Descartes, 75231 Paris, France}

\begin{abstract}
We introduce wavefront shaping as a tool for optimizing the sensitivity in nano-optomechanical measurement schemes. We perform multimode output analysis of an optomechanical system consisting of a focused laser beam coupled to the transverse motion of a tapered cantilever, and demonstrate that wavefront shaping enables a 350-fold enhancement of the measurement signal-to-noise ($+25.5\,\mathrm{dB}$) compared to standard split-detection, close to the quantum precision limit. Our results open new perspectives in terms of sensitivity and control of the optomechanical interaction.
\end{abstract}

\date{\today}

\maketitle

\section*{Introduction}
Optomechanics investigates the interactions between electromagnetic and mechanical degrees of freedom \cite{aspelmeyer2014cavity}. In just $30$ years, the field has made important progress, including the demonstration of ground-state cooling \cite{chan2011laser,delic2020cooling}, quantum correlations \cite{purdy2013observation,militaru2022ponderomotive,magrini2022squeezed} and remote micromechanical entanglement \cite{riedinger2018remote}. These milestones crucially rely on the concept of nano-optomechanical systems, that are devices harnessing the enhanced sensitivity of strongly confined electromagnetic degrees of freedom to mechanical perturbations, and vice versa \cite{li2008harnessing,eichenfield2009picogram,anetsberger2009near}.

Despite its remarkable effectiveness, this approach remains challenging to develop, which notably stems from the increased susceptibility of nanoscale-confined optical fields to boundary conditions. Thus, the design and optimization of nano-optomechanical systems essentially rely on advanced numerical simulations aiming at maximizing the optomechanical coupling, within restricted illumination conditions. The influence of the input state is however critical to measurement precision in general, a fact that is well established in quantum estimation theory \cite{helstrom1969quantum} and which was recently highlighted in the context of coherent scattering measurements \cite{bouchet2021maximum,gutierrez2024reaching}.

\begin{figure}[tb]
\includegraphics[width=85mm]{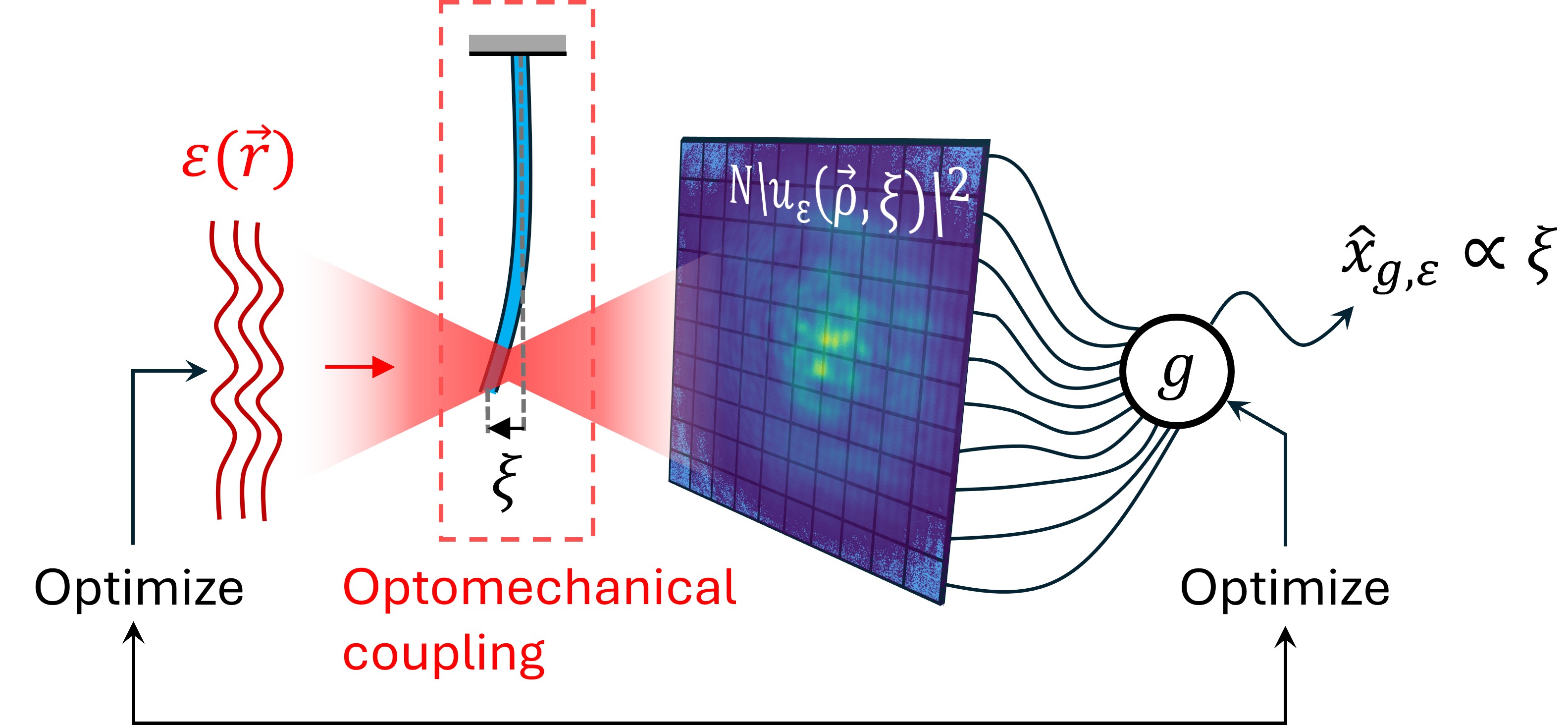} \centering 
\caption{\textbf{Principle of the experiment.} The tip's lateral displacement $\xi$ of a suspended nanocantilever is measured via the modifications its displacement produces on the intensity distribution of the transmitted light. Both the output estimator and input mode are optimized in order to maximize the measurement's performances.}
\label{Fig1}%
\end{figure}

In this work, we optimize the optical probe shape to enhance the sensitivity of a nano-optomechanical scheme, consisting of a single-pass focused laser beam coupled to the lateral displacement of a tapered cantilever (see Fig. \ref{Fig1}). Based on a multimode output analysis of our system, we demonstrate that wavefront shaping enables to bring the optomechanical measurement close to the quantum precision limit \cite{helstrom1969quantum}. This represents a $350$-fold sensitivity enhancement over the $\mathrm{TEM}_{00}$-driven split detection scheme commonly used for ultra-sensitive nano-optomechanical detection \cite{sanii2010high,arcizet2011single,huang2011direct,jain2016direct}. Our results suggest that wavefront shaping optimization may strongly boost the sensitivity of a wide class of nano-optomechanical experiments currently under development. 

\section*{Experimental setup}
\begin{figure}[t!]
\includegraphics[width=85mm]{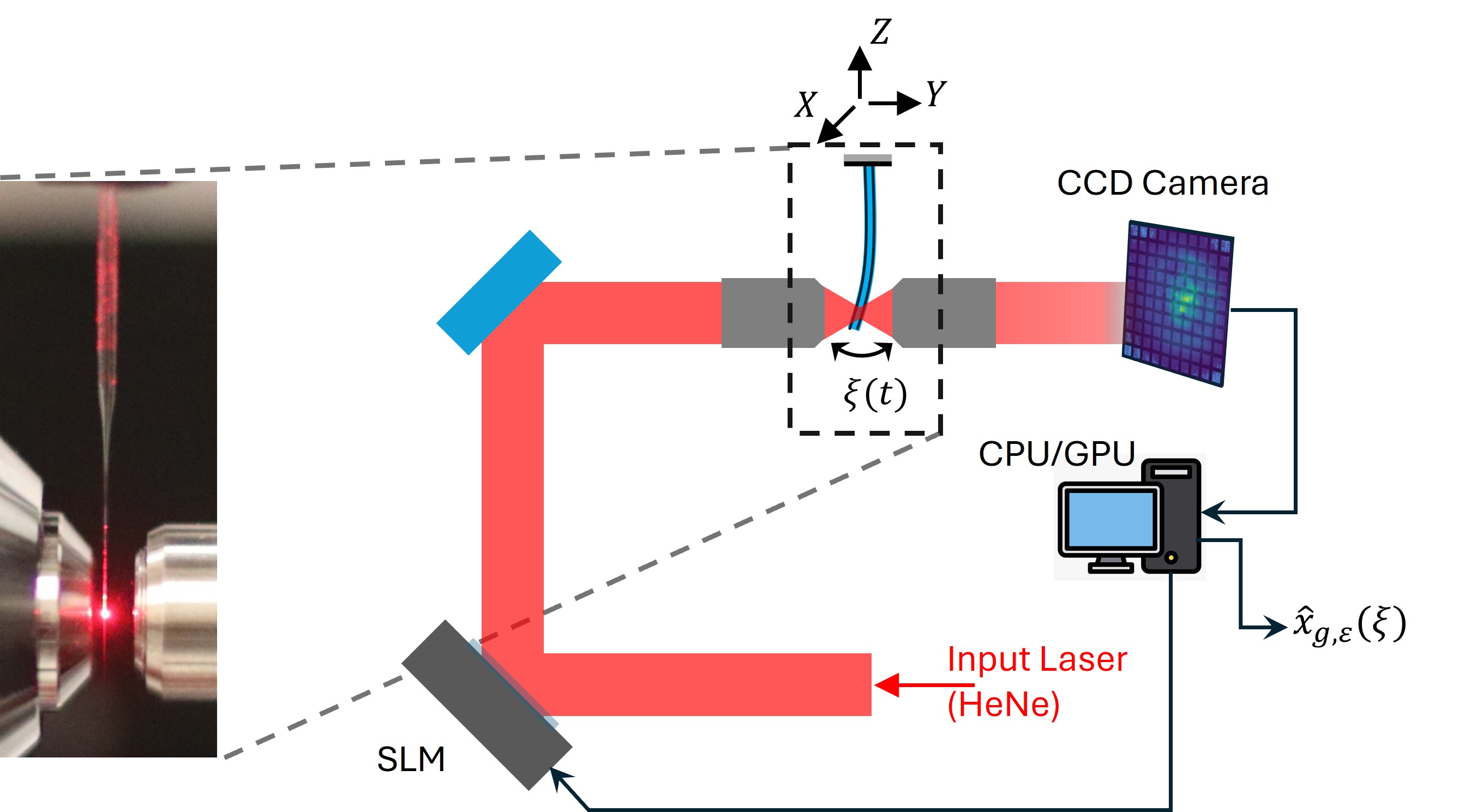} \centering
\caption{\textbf{Schematic of the experimental setup.} A single-mode $632\,\mathrm{nm}$ He-Ne laser is focused on a nano-optomechanical system consisting of a suspended tapered borosilicate capillary. A reflective liquid crystal spatial light modulator is inserted on the He-Ne optical path, enabling input wave-front shaping. The transmitted light is collected by a CMOS sensor and further analyzed using a computer station. Inset: Photograph of the optomechanical system placed at the focus of the input objective (on the left side of the picture).}%
\label{Fig2}%
\end{figure}

A schematic of our experimental setup is shown in Fig.~\ref{Fig2}. A single-mode HeNe laser is expanded onto the surface of a liquid crystal on silicon spatial light modulator (LCOS-SLM X15213, Hamamatsu), whose surface is imaged on the pupil of a $10\times$ microscope objective. The optomechanical system is mounted onto a 3-axis nanopositioning stage enabling to precisely adjust the capillary's tip position in the focal plane of the microscope objective. The optomechanical device consists of a borosilicate tapered capillary (see inset in Fig. \ref{Fig2}), whose external diameter typically decreases from $1\,\mathrm{mm}$ down to a few hundreds of $\mathrm{nm}$, from one extremity to the other \cite{antoni2023detection}.  The light scattered by the nano-optomechanical device is collected in transmission in the front focal plane of a high-numerical aperture aspherical lens ($\mathrm{NA=0.55}$), and further analyzed by means of a CMOS camera. The camera and SLM are linked to a computing station used to optimize both the output and input measurement modes, so as to maximize the optomechanical sensitivity. The mechanical degree of freedom of interest in this study is hosted by a flexural mode with mechanical resonance frequency $\Omega_{\mathrm{m}}/2\pi=133\,\mathrm{Hz}$, which enables the construction of a real-time motion estimator from the frame data acquired with the CMOS camera. Given the $\mathrm{cm}-$scale spatial footprint of this mechanical mode, we choose to adjust the vertical position of the optomechanical device a few tens of $\mu\mathrm{m}$ above its apex, so as to illuminate a capillary cross-section larger than the optical waist, thereby ensuring maximal optomechanical overlap. In the following, the mechanical displacement is resonantly driven using acoustic waves generated by a standard laptop speaker.

\section*{Optomechanical motion estimation}
The linear motion signal is defined as:
\begin{eqnarray}
\hat{x}_{g,\varepsilon}(t)&=&N\int\mathrm{d}^2\vec{\rho}\,g(\vec{\rho})\left(|u_{\varepsilon}(\vec{\rho},\xi(t))|^2-|u_{\varepsilon}(\vec{\rho},\xi=0)|^2\right),\nonumber\\
\label{eq:1}
\end{eqnarray}

 Here $\varepsilon$ denotes the input field, $u_\varepsilon(\vec{\rho},\xi)$ the (input-dependent) complex amplitude of the transmitted field at the pixel localized at position $\vec{\rho}$, $\xi$ the time-dependent tip displacement at the optomechanical interaction region, $g$ the pixel gain distribution, and $N$ the average number of photons accumulated during the acquisition of a single frame ($N=7\times10^4$ in this work). In practice, Eq. \ref{eq:1} means that the value of the motion signal at time $t$ is obtained as the overlap between a suitably defined pixel gain function and the difference between the current intensity frame $|u_{\varepsilon}(\vec{\rho},\xi(t))|^2$ and a reference frame $|u_{\varepsilon}(\vec{\rho},\xi=0)|^2$, recorded in absence of motion modulation.
 
 A better grasp of Eq. \ref{eq:1} can be gained by expanding it to first order in $\xi$. The output field then writes $|u_{\varepsilon}(\vec{\rho},\xi)|\simeq|u_{\varepsilon}(\vec{\rho},\xi=0)|+\frac{\xi}{a_\varepsilon}v_\varepsilon(\vec{\rho})$, with $a_\varepsilon        =\sqrt{1/\int\mathrm{d}^2\vec{\rho}\left(\frac{\mathrm{d}|u_\varepsilon|}{\mathrm{d}\xi}\right)_{\xi=0}^2}$ a characteristic length further referred to as the 'optomechanical waist', and $v_\varepsilon$ the normalized, first-order expansion of the transmitted field amplitude, yielding to:

 \begin{eqnarray}
 \hat{x}_{g,\varepsilon}(t)&\simeq&2N\frac{\xi(t)}{a_\varepsilon}\int\mathrm{d}^2\vec{\rho}\,g(\vec{\rho})|u_{\varepsilon}(\vec{\rho},\xi=0)|v_\varepsilon(\vec{\rho}).\label{eq:2}
 \end{eqnarray}

Besides being explicitly proportional to the tip displacement $\xi$ (which is the least requirement from a linear measurement), Eq. \ref{eq:2} shows that the signal sensitivity is essentially set by two factors. On the one hand, it is inversely proportional to the optomechanical waist, which is determined by the input field $\varepsilon$. On the other hand, the integral term in Eq. \ref{eq:2} defines an inner product in the camera's plan, therefore measuring the projection $\langle v_{g,\varepsilon},v_\varepsilon\rangle$ of a certain gain-dependent measurement mode $v_{g,\varepsilon}=g(\vec{\rho})|u_\varepsilon(\vec{\rho},\xi=0)|$ over $v_\varepsilon$, which can be viewed as a reference mode. In particular, the sensitivity of the measurement is maximized when the measurement and reference modes are collinear $v_{g,\varepsilon}\propto v_\varepsilon$, resulting in the optimal gain condition $g_\varepsilon(\vec{\rho})\propto v_\varepsilon(\vec{\rho})/|u_\varepsilon(\vec{\rho},\xi=0)|$. Note that this expression of the optimal gain is analogous to the Wigner-Smith limit \cite{wigner1955lower,smith1960lifetime,ambichl2017focusing}, allowing to reach the Cramér-Rao bound (CRB), which maximizes the Fisher information available to the measurement of $\xi$ for a given coherent input state $\varepsilon$ \cite{delaubert2007quantum,gutierrez2024reaching}. Here, we use the gain convention ensuring a unit norm for the associated detection mode, $\langle v_{g,\varepsilon}, v_{g,\varepsilon}\rangle=1$. Within this convention, $\langle v_{g,\varepsilon},v_\varepsilon\rangle\leq 1\,\forall g$, the equality being reach for the CRB only, $g=g_\varepsilon$.

In this work, we thus propose to investigate those two factors impacting the measurement sensitivity, that are the camera pixels gain and the incident measurement field, respectively.

\section*{Measurement mode optimization}
We first propose to test the output mode optimization by trying various pixel gain functions $g(\vec{\rho})$. The microscope objective is first fed by a $\mathrm{TEM}_{00}$ mode, the SLM acting as a simple mirror. A thousand frames are recorded from the CMOS camera while driving the mechanical motion, to which a reference frame is subsequently subtracted. Each of the resulting arrays is multiplied component-wise by the chosen pixel gain function, the corresponding value of the motion signal being obtained by summing all its elements (see Eq. \ref{eq:1}).

At first glance, the mechanical motion is observed to result in an overall lateral translation of the intensity distribution measured on the CMOS. This is reminiscent of a 'beam displacement' coupling, whereby the optomechanical interaction results in an effective displacement of the output optical axis, relative to the input mode's \cite{treps2002surpassing}. This behavior has been observed on a wide variety of nano-optomechanical coupling platforms, which has largely justified the use of balanced-type detectors \cite{sanii2010high,huang2011direct,arcizet2011single}: Indeed, for a pure coherent $\mathrm{TEM_{00}}$ beam displacement measurement, it can be shown that the measurement mode associated with a split-detector (also known as 'flipped mode') performs close to the CRB, with $\langle v_{g,\varepsilon},v_\varepsilon\rangle=\sqrt{2/\pi}\simeq0.8$ \cite{fabre2000quantum,treps2002surpassing,delaubert2007quantum}. Concurrently, the optimal detection mode for such measurement $v_\varepsilon\propto \frac{\mathrm{d}|u_\varepsilon|}{\mathrm{d}\xi}\propto\left(\vec{\rho}\cdot\vec{e_\xi}\right)|u_\varepsilon(\vec{\rho},\xi=0)|$, with $\vec{e_\xi}$ the direction of the mechanical motion. Note that the associated motion signal $\propto \int\mathrm{d}^2\vec{\rho}\left(\vec{\rho}\cdot\vec{e_\xi}\right)|u_\varepsilon(\vec{\rho},\xi(t))|^2$ formally amounts to calculating the barycenter of the frame, which can also be viewed as tracking its 'center'.

\begin{figure}[t!]
\includegraphics[width=85mm]{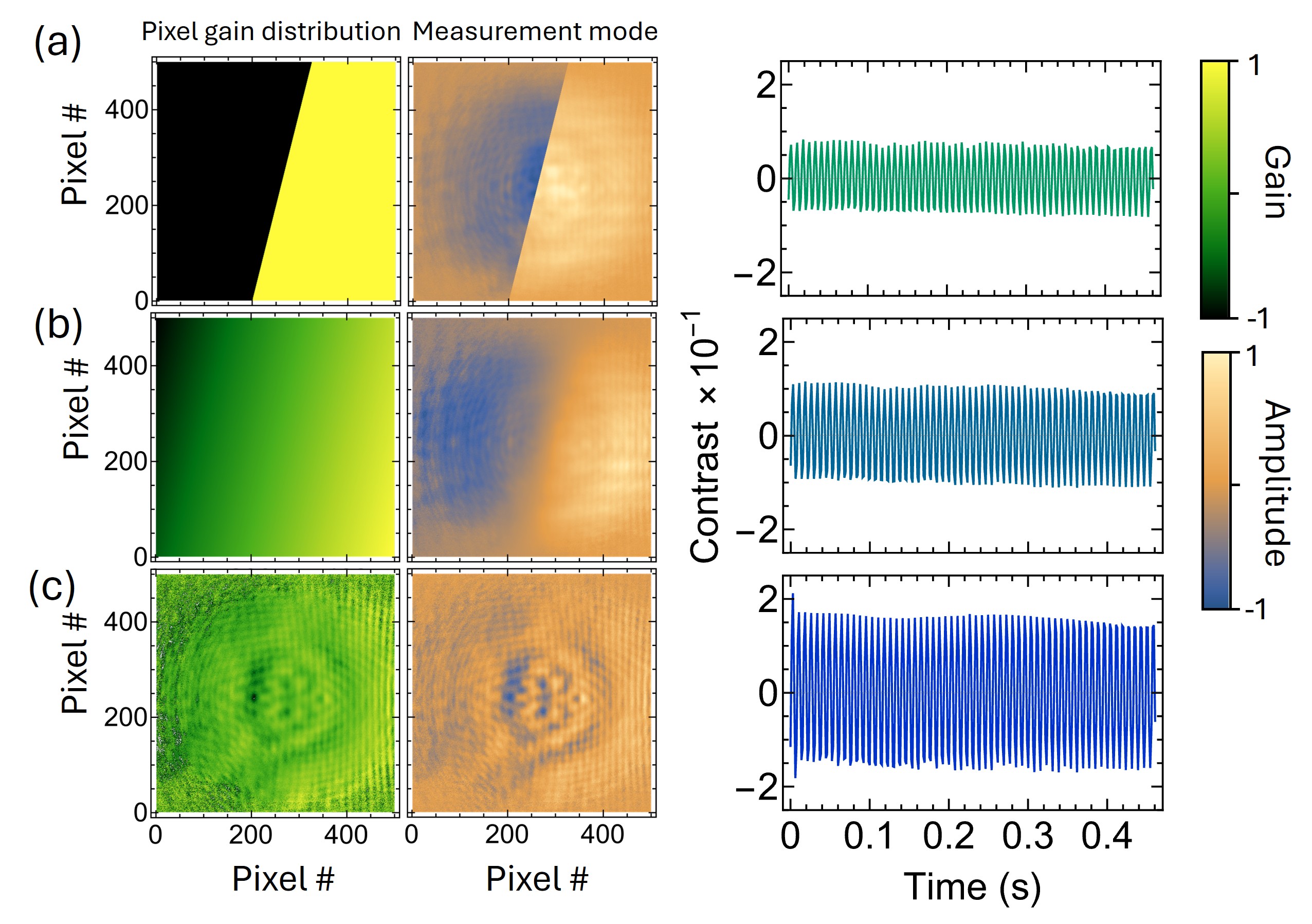} \centering
\caption{\textbf{Output mode optimization.} Each row shows the pixel gain distribution $g_{i}$ (left), the associated measurement mode $v_{g,\varepsilon}$ and the time evolution of the corresponding motion signal $\hat{x}_{g,\varepsilon}$, in the case of $\varepsilon_{00}=\mathrm{TEM}_{00}$ input illumination. (a) Split signal. (b) Tracking signal. (c) Optimal signal.}%
\label{Fig3}%
\end{figure}

\begin{figure*}[t!]
\includegraphics[width=17cm]{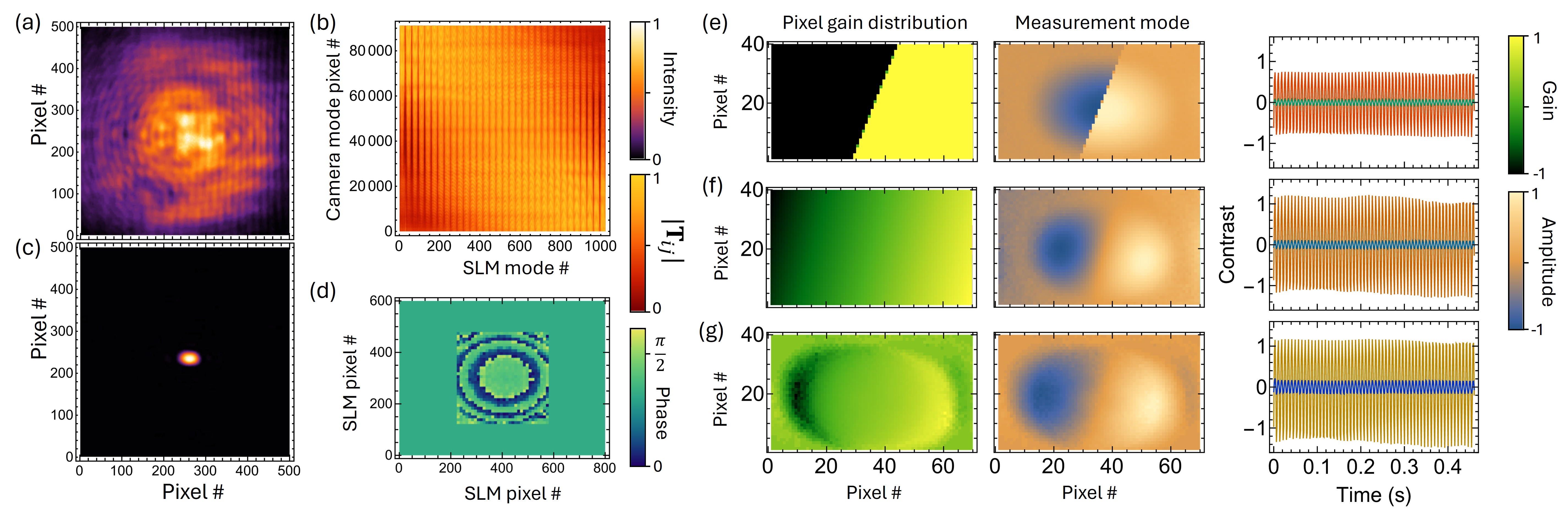} \centering
\caption{\textbf{Input-output sensitivity optimization.} (a) Unshaped transmitted output field intensity distribution $u_{00}$. (b) Norm of the optical transmission matrix, expressed in the SLM-CMOS pixels' bases. (c) Shaped transmitted output field intensity distribution. (d) Phase mask applied to the SLM in the shaped configuration. (e) Pixel gain distribution (left), measurement mode and time evolution for the split signal ($\varepsilon=\varepsilon_{\mathrm{foc}}$). The small, centered light-color trace represents the signal obtained with $\mathrm{TEM}_{00}$ input illumination.  (f,g) same as (e) for the tracking and optimal signals, respectively.}  
\label{Fig4}
\end{figure*}

This argument justifies our choice to specifically focus our attention on both the 'split' and 'tracking' gains, respectively defined (up to a multiplicative constant) as $g_s(\vec{\rho})=-1$ if $\vec{e}_{\xi}\cdot\vec{\rho}>0$ and $+1$ otherwise on the one hand; and $g_t(\vec{\rho})=\left(\vec{\rho}\cdot\vec{e_\xi}\right)$ on the other hand. These pixel gain distributions are shown in Fig. \ref{Fig3} (a--b), along with the corresponding measurement modes. Note the tilted separation of the split gain, reflecting the non-perfectly horizontal motion of the capillary's tip, whose direction was determined from the 2-dimensional trajectory of the tracking signal. Additionally, Fig. \ref{Fig3}(c) shows the optimal pixel gain distribution $g_\varepsilon(\vec{\rho})=v_\varepsilon(\vec{\rho})/|u_\varepsilon(\vec{\rho},\xi=0)|$, together with the associated optimal measurement mode $v_\varepsilon$ for the input field $\varepsilon$. At this stage, two observations can already be made. First, the split and tracking modes seem to be roughly similar to those associated with $\mathrm{TEM}_{00}$ beam displacement, which stems from the general appearance of the diffraction spot. Secondly and despite the $+/-$ anti-symmetry remaining visible, the optimal measurement mode markedly reveals small scale variations that were not present otherwise. This anticipates the availability of a significant improvement of the measurement sensitivity through pixel gain optimization. 

The performance of each gain configuration is further quantified by plotting the time-evolution of the mechanically-induced optical contrast, defined as the motion signal divided by the single-frame photon number (right panel in Fig. \ref{Fig3}): In particular, the amplitude of each curve corresponds to the optomechanical modulation depth $\mu_{g,\varepsilon}=\frac{\sqrt{2}\Delta  \hat{x}_{g,\varepsilon}}{N}$. As anticipated above, a clear sensitivity improvement is observed moving from the split, to tracking and then optimal pixel gain distributions, with corresponding modulation depths respectively evaluating to $\mu_{s,00}\simeq 0.08$, $\mu_{t,00}\simeq 0.11$ and $\mu_{00}\simeq 0.17$ (the index $00$ stemming for $\varepsilon_{00}\equiv\mathrm{TEM}_{00}$). Interestingly, the increase in sensitivity from the split to tracking configuration is $\simeq 2.9\,\mathrm{dB}$, that is more than the $\simeq 2\,\mathrm{dB}$ expected for $\mathrm{TEM}_{00}$ beam displacement measurement \cite{delaubert2007quantum}. Furthermore, the optimal pixel gain distribution enables a $\simeq4.1\mathrm{dB}$  signal enhancement compared to tracking detection: This reflects that an important fraction of the input modes couples to mechanical motion via processes other than beam displacement. 

\section*{Wavefront shaping-enhanced optomechanical coupling} 

We subsequently turn to optimizing the input field, which prominently affects the measurement sensitivity, notably via the optomechanical waist $a_\varepsilon$, which measures the responsiveness of the output field with respect to mechanical motion. In this work, we follow a deterministic approach and choose to specifically concentrate on reducing the optomechanical waist by focusing the field transmitted by the optomechanical device. This again justifies in analogy with beam displacement measurement, whose sensitivity is fully controlled by the confinement of the interaction gradient \cite{delaubert2007quantum,delaubert2008quantum}.

We perform output focusing following the phase conjugation method described in \cite{popoff2010measuring}, which is essentially two steps. First, the optical transmission matrix $\mathbf{T}$ is determined by sequential loading of the Hadamard basis vectors on the SLM, while the capillary is at rest. Second, this matrix is used to determine the phase mask to be applied on the SLM in order to complete focusing through the optical medium. To do so, we define a target spot $u_{\mathrm{foc}}$, whose size is adjusted to match an 'optical grain' (in analogy with a speckle grain), defined from the spatial autocorrelation of the output field intensity $|u_{00}(\vec{\rho},\xi=0)|^2$ resulting from the transmission of the $\mathrm{TEM}_{00}$ input mode. The SLM is subsequently addressed with a mask such that the input wavefront be shaped to $\varepsilon_{\mathrm{foc}}=\mathbf{T}^{\dagger}u_{\mathrm{foc}}$ ($\dagger$ standing for conjugate transpose), whose transmission through the capillary is sensitive to the time reversal operator $\mathbf{T}\mathbf{T}^{\dagger}$ , therefore completing phase conjugation.

Figures \ref{Fig4}(a--c) orderingly show the intensity distribution of the output field $|u_{00}(\vec{\rho},\xi=0)|^2$, the optical transmission matrix expressed in the SLM-CMOS pixels' bases, and the intensity distribution of the focused output field $|u_{\mathrm{foc}}(\vec{\rho},\xi=0)|^2$. The corresponding phase mask applied to the SLM is shown in Fig. \ref{Fig4}(d). We subsequently proceed with an analysis similar to that outlined in Fig. \ref{Fig3}, by computing $3$ motion signals associated with the pixel gain distributions $g_s$, $g_t$ and $g_{\mathrm{foc}}$. The results are shown in Figures \ref{Fig4}(e--g). A clear improvement of the sensitivity is observed compared to what was obtained with $\mathrm{TEM}_{00}$ input illumination, with the new modulation depths evaluating to $\mu_{s,\mathrm{foc}}\simeq0.9$ ($\mu_{s,\mathrm{foc}}/\mu_{s,00}\simeq11.3$, $\sim\,+21.1\,\mathrm{dB}$ sensitivity), $\mu_{t,\mathrm{foc}}\simeq1.3$ ($\mu_{t,\mathrm{foc}}/\mu_{t,00}\simeq12$, $\sim\,+21.6\,\mathrm{dB}$) and $\mu_{\mathrm{foc}}\simeq1.4$ ($\mu_{\mathrm{foc}}/\mu_{00}\simeq12$, $\sim\,\,+9.2\,\mathrm{dB}$). This sensitivity enhancement mainly relies on the significant decrease of the optomechanical waist, which alone contributes to $\sim\,+14.8\,\mathrm{dB}$,  with $a_{00}/a_{\mathrm{foc}}\simeq5.5$. The additional gain observed in the split configuration is consistent with an improved coupling of the shaped input mode to beam displacement, which is further confirmed by the performance of the tracking signal, very close to the optimal's: From those results, we retain that output beam focusing enables us to both enhance motion sensitivity, and homogenize the nature of the optomechanical coupling to quasi-pure beam displacement.

\section*{Discussion}
\emph{Optomechanical nonlinearities.} Note that the large modulation depths reported above challenge our linear optomechanical transduction hypothesis (used for deriving Eq. \ref{eq:2}), which reflects in the values of $\mu_{t,\mathrm{foc}}$ and $\mu_{\mathrm{foc}}$, both $>1$. The presence of nonlinearities is confirmed by the coupling of the output mode intensity to mechanical motion ($\sim 10\%$ modulation depth, not shown), which was not observed under $\mathrm{TEM}_{00}$ illumination. These nonlinearities are the consequence of a motion amplitude that has become large compared to the optomechanical waist ($\xi\gtrsim a_{\mathrm{foc}}$), and which typically introduce a bias over the determination of the measurement mode, explaining why $\mu_{t,\mathrm{foc}}$ and $\mu_{\mathrm{foc}}$ are observed to overshoot above unity.
\begin{figure}[htpb]
\includegraphics[width=85mm]{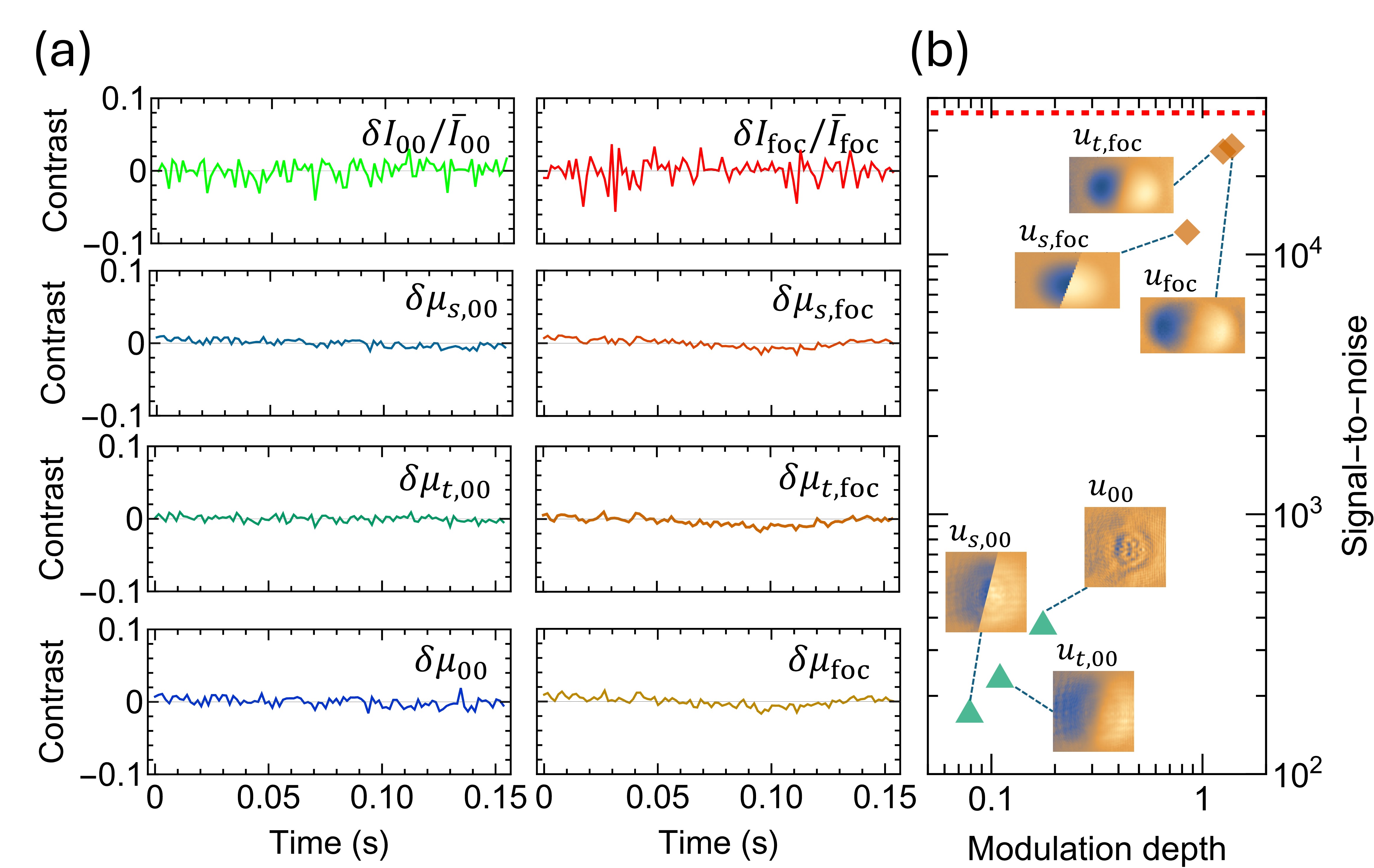} \centering
\caption{\textbf{Measurement noises and  signal-to-noise ratios.} (a) Noises for various pixel gain distributions, both in the unshaped (left) and shaped (right) configurations. From top to bottom are shown the intensity, the split mode, the tracking mode and the optimal mode noises. (b) Signal-to-noise ratio associated with the various measurement gains, both in the unshaped (triangles) and shaped (diamonds) configurations. The red, dashed line represents the quantum precision limit.}%
\label{Fig5}%
\end{figure}

\emph{Measurement noise sensitivity.} Besides the magnitude of their responses toward the parameter to be measured, it is also essential to address the measurement noises.  We do so by acquiring the output intensity fluctuations while the optomechanical capillary resting at a fixed position. The time evolutions of the corresponding signals are subsequently computed using Eq. \ref{eq:1}. The results are shown in Fig. \ref{Fig5}(a). The rows from top to bottom represent the output intensity noise, followed by the split, tracking, and optimal signal noises, both in the unshaped ($u_{00}$) and shaped ($u_{\mathrm{foc}}$) configurations (left and right columns, respectively). One immediately notes the presence of a sizable classical intensity noise ($\sim10\,\mathrm{dB}$ above the split signal noise), emphasizing that the role of measurement shaping does not limit itself to maximizing motion transduction, but may also enact efficient noise eating. Each trace is used to determine the associated measurement equivalent noise variance $\Delta\mu_{g,\epsilon}^2$, which is compared to the motion-induced modulation depth so as to form the signal-to-noise ratio (SNR), $\mathrm{SNR}_{g,\epsilon}=\mu_{g,\epsilon}^2/2\Delta\mu_{g,\epsilon}^2$. The results are summarized in Fig. \ref{Fig5}(b). A large increase of the SNR is observed to the benefit of the shaped vs. unshaped configuration, which essentially reflects similar levels of noise in both cases. Moreover, the SNR generally increases at higher signal sensitivity, with very similar values being reached with the tracking and optimal pixel gains. This further confirms that the shaped input is almost perfectly coupled to the capillary via beam-displacement. Finally, we see that a maximum $\mathrm{SNR}_{\mathrm{foc}}\simeq 2.6\times10^4$ is reached with the focused optimal pixel gain distribution, just $1.3\,\mathrm{dB}$ away from the quantum precision limit $\mathrm{SNR}_{\mathrm{q}}=N/2\simeq3.5\times10^4$ \cite{helstrom1969quantum,paris2009quantum}. We interpret this discrepancy as yet another consequence of the coupling nonlinearities, which are responsible both for an imprecision in the determination of the CRB mode, and for a clipping effect resulting in a reinforcement of the noise sensitivity, to the detriment of the signal.

\emph{Perspectives} As outlined above, the sampling limitations of our imaging system are the main reason why we have limited ourselves to working with a low-frequency mechanical mode. Extending the present study to higher frequency domains will require fast multimode imaging output devices, which may include multi-plane light converters \cite{labroille2014efficient,rouviere2024ultra,choi2024quantum} and fast cameras. Higher optomechanical sampling rate may also be achieved by means of an heterodyne detection scheme \cite{popoff2011exploiting}, where optimizing the measurement projection would be made by shaping the reference arm \cite{bouchet2021maximum}. Additionally, the fact that we are able to reach the quantum precision limit in pure intensity measurement settings represents a particular case, which essentially relates to the structure of the particular optomechanical interaction of interest, resulting in a saturation of the quantum precision limit by the tracking gain distribution. In general, both phase and amplitude may couple to mechanical motion, in which case field-sensitive imaging must be used \cite{delaubert2007quantum}.

\section*{Conclusion}
In conclusion, we have reported a new type of nano-optomechanical experiment, whereby input-output wavefront shaping enables to optimize the measurement sensitivity close to the quantum precision limit. By focusing the output measurement mode, we show a $+25.5\,\mathrm{dB}$ sensitivity enhancement starting from a standard $\mathrm{TEM}_{00}$ input/split-detector output scheme, scaling the quantum detection efficiency from $\sim0.2\%$ up to $\sim74\%$. This result can be put into perspective with recent split detector-based transverse optomechanical detection experiments, whose quantum efficiencies are expected to peak around $10\,\%$ \cite{tebbenjohanns2019optimal}. In our case, the effect of input mode shaping is to address the optomechanical system with a maximal eigenstate for the beam-displacement operator, whose quantum precision limit is available to pure intensity measurement.

Our results have the potential to be extended to all types of optomechanical coupling, which would generally require field-sensitive real-time multimode output analysis \cite{delaubert2007quantum}. Additionally, wavefront shaping may also serve to tailor measurement backaction \cite{orazbayev2024wave}, and to suppress the sensitivity towards unwanted degrees of freedom that may affect coherent processes \cite{saarinen2023laser}. In particular, nanosystems currently faced with major thermalization issues, including quantum cryogenic optomechanical systems \cite{ren2020two,fogliano2021ultrasensitive}, may be among those benefiting the most from our approach. Indeed, the ability of selectively channeling the probe energy in the degrees of freedom of interest represents a net increase of the optomechanical coupling rate, which generally allows a sizeable power reduction (compared to the unshaped scenario) while maintaining equivalent sensitivities.

\bibliography{bib}
\end{document}

\title{Supplementary Information: Wavefront shaping enhanced nano-optomechanics down to the quantum precision limit}

\author{A. G. Tavernarakis\,\orcidlink{0000-0002-9780-2514}}
\affiliation{Université Paris-Saclay, CNRS, ENS Paris-Saclay, CentraleSupélec, LuMIn, 91405, Orsay,
France}

\author{R. Gutiérrez-Cuevas\,\orcidlink{0000-0002-3451-6684}}
\affiliation{Institut Langevin, ESPCI Paris, Université PSL, CNRS, 75005, Paris, France}
\affiliation{Université Paris-Saclay, CNRS, ENS Paris-Saclay, CentraleSupélec, LuMIn, 91405, Orsay,
France}

\author{L. Rondin\,\orcidlink{0000-0002-4833-2886}}
\affiliation{Université Paris-Saclay, CNRS, ENS Paris-Saclay, CentraleSupélec, LuMIn, 91405, Orsay,
France}

\author{T. Antoni\,\orcidlink{0000-0002-0697-0986}}
\affiliation{Université Paris-Saclay, CNRS, ENS Paris-Saclay, CentraleSupélec, LuMIn, 91405, Orsay,
France}

\author{S. M. Popoff\,\orcidlink{0000-0002-7199-9814}}
\affiliation{Institut Langevin, ESPCI Paris, Université PSL, CNRS, 75005, Paris, France}

\author{P. Verlot\,\orcidlink{0000-0002-5105-3319}}
\email{pierre.verlot@universite-paris-saclay.fr}
\affiliation{Université Paris-Saclay, CNRS, ENS Paris-Saclay, CentraleSupélec, LuMIn, 91405, Orsay,
France}
\affiliation{Institut Universitaire de France, 1 rue Descartes, 75231 Paris, France}

\date{\today}

\pacs{42.50.-p, 03.65.Ta, 42.50.Lc}
\maketitle

\section{Beam displacement measurement}
This section outlines the paradigm of multipixel optomechanical beam displacement measurement \cite{fabre2000quantum,treps2002surpassing,delaubert2007quantum}, and notably justifies the prominence of the tracking estimator in this particular context. The input mode is assumed to be a $\mathrm{TEM}_{00}$ with waist $w_0$ and wave vector $\vec{k}=k\vec{e}_z$, given by:

\begin{figure}[hbt]
\includegraphics[width=85mm]{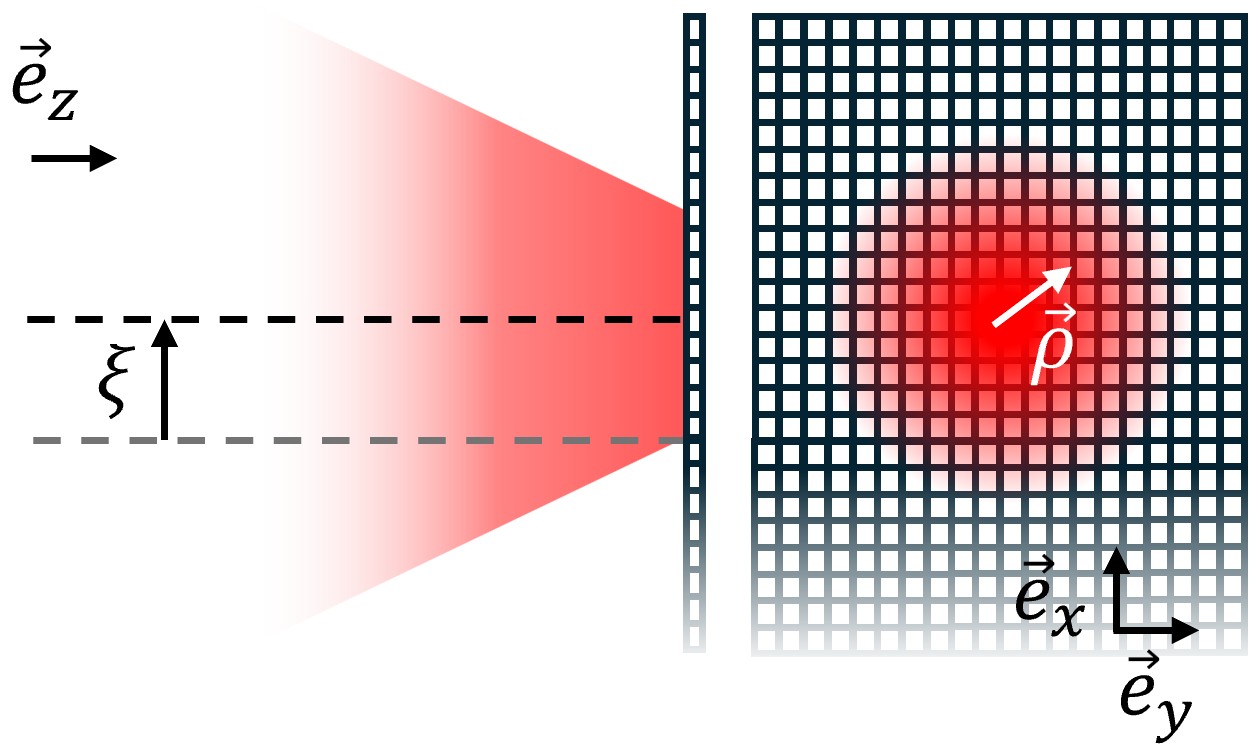} \centering
\caption{\textbf{Beam displacement measurement.} A $\mathrm{TEM}_{00}$ beam is assumed to experience optomechanically-induced beam displacement, resulting in a transverse shift of its optical axis. The resulting space modification of the intensity distribution $|u(x+\xi,y,z)|^2$ is detected at its focus on a multipixel detector.}%
\label{FigS1}%
\end{figure}

\begin{eqnarray}
\varepsilon_{00}(x,y,z)&=&\frac{1}{w(z)\sqrt{2\pi}}e^{-\frac{x^2+y^2}{w^2(z)}}e^{-ik\frac{x^2+y^2}{2R(z)}}e^{-i\phi_G(z)},\label{eq:1}
\end{eqnarray}

with $(x,y,z)$ the 3 space coordinates ($\vec{e}_x$ the transverse horizontal, $\vec{e}_y$ the transverse vertical and $\vec{e}_z$ the propagation directions, respectively), $R(z)=z+z^2/z_R$ the radius of curvature ($z_R=kw_0^2/2$ the Rayleigh range), $w(z)=w_0\sqrt{1+\left(z/z_R\right)^2}$ the local beam size and $\phi_G(z)=\mathrm{arctan}\left(z/z_R\right)$ the Gouy phase. Beam displacement amounts to a simple lateral shift $\xi$ off the optical axis in the motion direction. Assuming the beam to be displaced in the transverse horizontal direction, the output mode is therefore given to first order by:

\begin{eqnarray}
u_{00}(x+\xi,y,z)&=&\varepsilon_{00}(x,y,z)+\xi\frac{\partial \varepsilon_{00}}{\partial x}(x,y,z)\nonumber\\
&=&\left(1+\xi\frac{2x}{w(z)}\right)\varepsilon_{00}(x,y,z).\label{eq:2}
\end{eqnarray}

Further assuming that the camera images the beam focus ($z=0$), the optimal detection mode $v_{00}\propto\frac{d|u_{00}|}{d\xi}$ is therefore given by:

\begin{eqnarray}
v_{00}(\rho_x,\rho_y)&\propto&2\rho_x \delta_{\mathrm{pix}}|\varepsilon_{00}(x,y,z=0)|,\label{eq:3}
\end{eqnarray}

where $\delta_{\mathrm{pix}}$ is a single pixel size, and  where the proportionality constant is determined such that $\langle v_{00},v_{00}\rangle=1$. Besides, the optimal pixel gain distribution $g_{00}(\vec{\rho})\propto v_{00}(\vec{\rho})/|u_{00}(\vec{\rho},\xi=0)|\propto\rho_x$, which identifies to the expression of the 'tracking' gain $g_t(\vec{\rho})\propto{\vec{\rho}\cdot\vec{e}_{\xi}}$ given in the main manuscript, assuming a purely horizontal motion ($\vec{e}_{\xi}=\vec{e}_x$). Eq. 1 from the main manuscript subsequently provides the expression of the associated motion estimator:

\begin{eqnarray}
\hat{x}_{g_t,00}&=&N\int\mathrm{d}^2\vec{\rho}\,\rho_x\left(|u_{00}(\vec{\rho},\xi)|^2-|u_{00}(\vec{\rho},\xi=0)|^2\right),\nonumber\\
&=&N\int\mathrm{d}^2\vec{\rho}\,\rho_x|u_{00}(\vec{\rho},\xi)|^2,\label{eq:4}
\end{eqnarray}
where we have used that $\rho_x\rightarrow\rho_x|u_{00}(\vec{\rho},\xi=0)|^2$ is an odd function, whose horizontal sum therefore cancels. Thus, Eq. \ref{eq:4} shows that the optimal estimator for measuring the displacement measurement of a $\mathrm{TEM}_{00}$ beam indeed amounts to evaluate the barycenter of its intensity distribution, as stated in the main text. 
\section{Mechanical motion direction}
\begin{figure}[hbt]
\includegraphics[width=85mm]{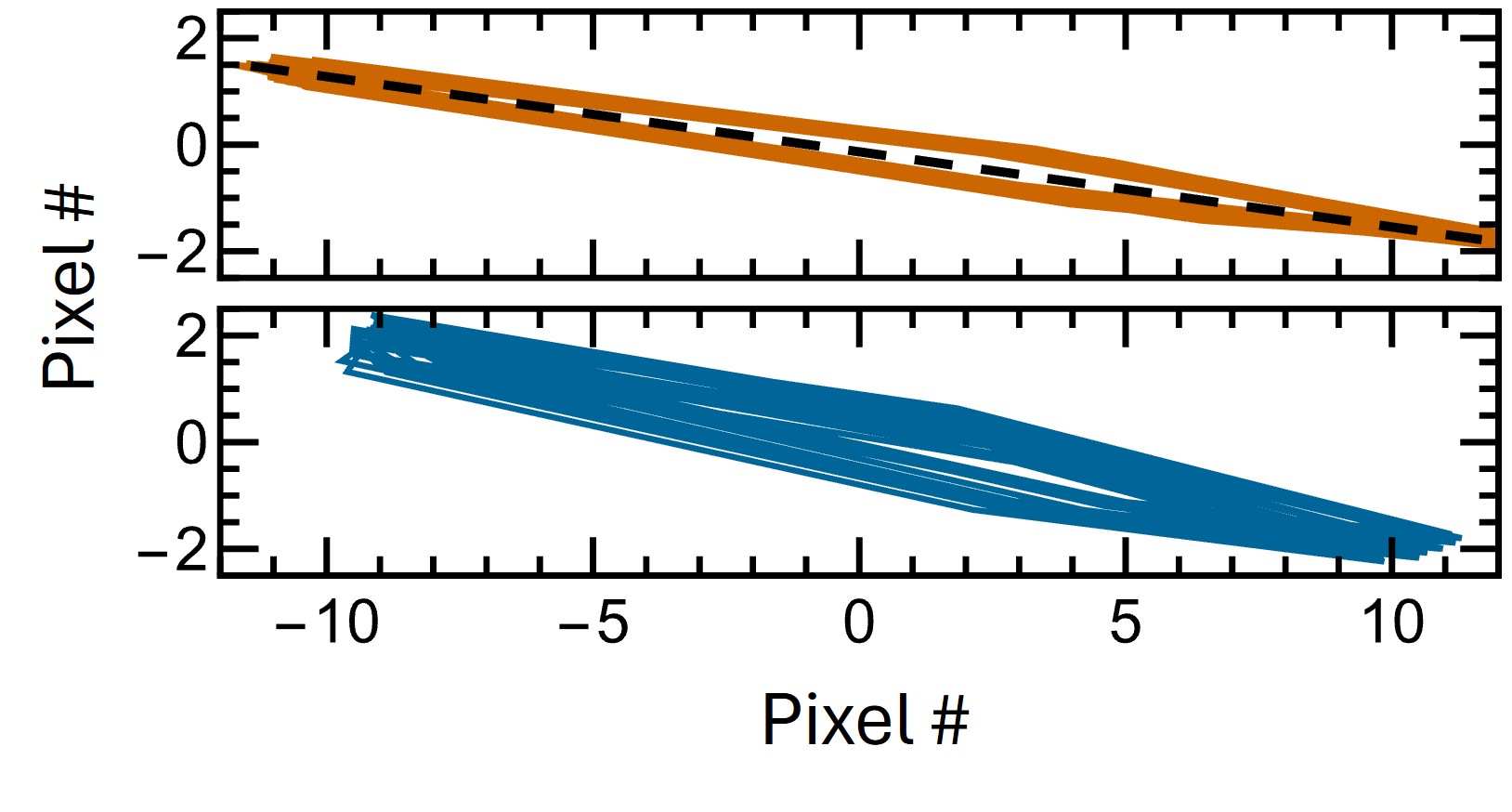} \centering
\caption{\textbf{2-dimensional tracking estimator trajectory.} Top: Trajectory obtained for with the shaped, 2d tracking estimator. Bottom: Same as top, for the unshaped 2d tracking estimator. }%
\label{FigS2}%
\end{figure}

To determine the actual direction of motion, we evaluate the tracking estimator both in the horizontal and vertical directions whilst the input field being optimized. The result is shown on Fig. \ref{FigS2} (top plot), where the tilted direction of motion clearly appears, and from which we infer the correct orientation of both the split and tracking pixel gain distributions (see Figs. 3 and 4 from the main manuscript). Also note the slightly elliptical shape of the trajectory, which indicates the presence of a small transduction effect in the orthogonal direction (further neglected throughout our work). The bottom plot represents the two-dimensional tracking estimator evaluated in the unoptimized input field configuration. While showing a similar behavior as that described above, one clearly sees that the corresponding trajectory is significantly noisier, and that the average motion direction differs from that inferred with the optimized input mode. This is reminiscent from the fact that the unoptimized input mode does not purely couple to beam displacement (also see main text for a more in-depth discussion). 

\section{Nonlinear motion-intensity coupling}

\begin{figure}[hbt]
\includegraphics[width=85mm]{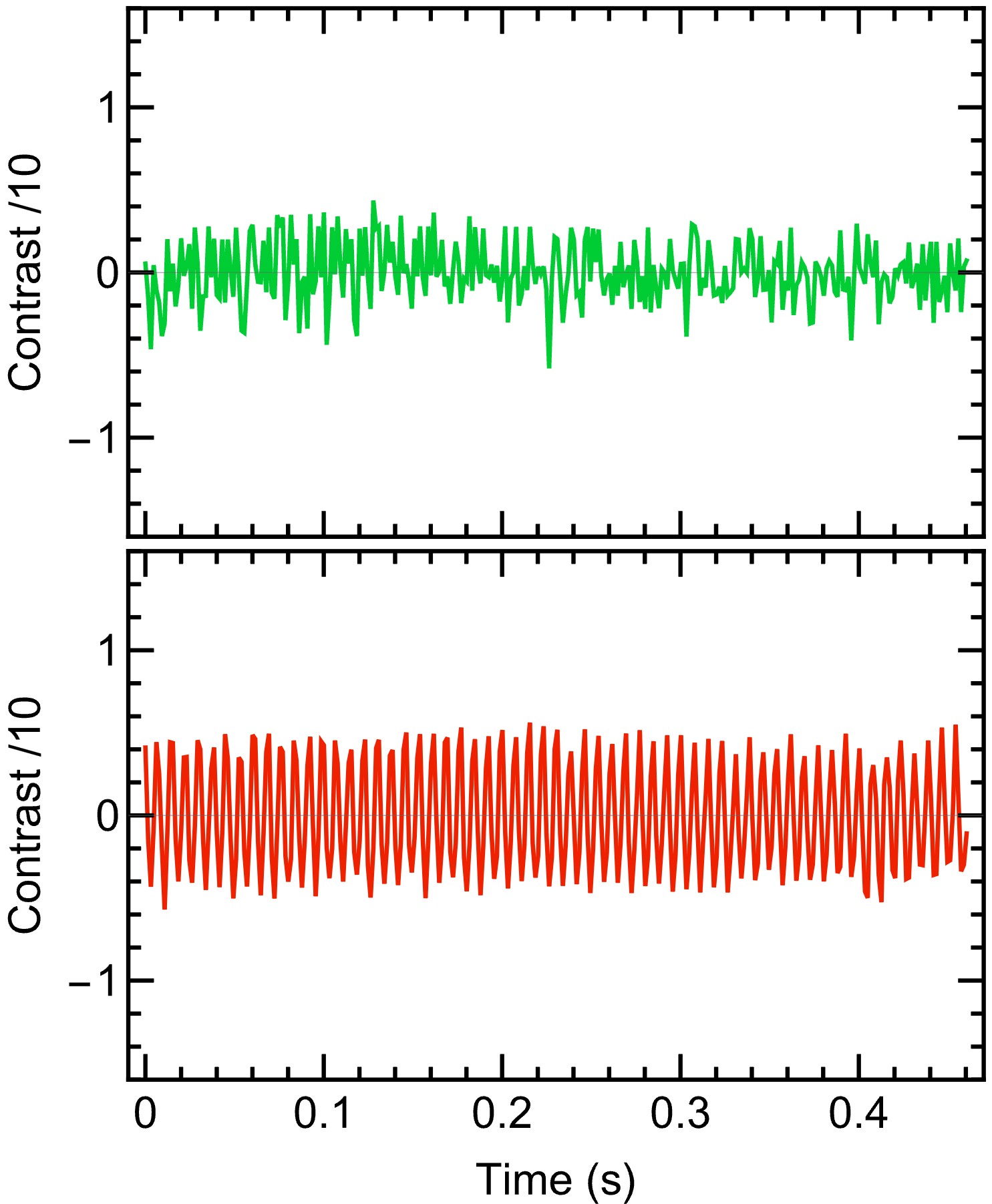} \centering
\caption{\textbf{Non-linear optomechanical coupling.} Top: Normalized intensity noise obtained as the time evolution of the sum of all pixels, in the unshaped configuration. Bottom: Same in the wavefront shaped configuration.}%
\label{FigS3}%
\end{figure}

As explained in the main manuscript, the large sensitivity enhancement enabled by wavefront shaping is associated with a drastic reduction of the optomechanical waist, which generally increases the sensitivity towards nonlinearities. These may notably manifest through a coupling between the mechanical motion and the transmitted intensity (obtained by summing all the pixels), which we assumed to be independent in the linear limit (see Eqs. 1 and 2 from the main manuscript, where the single-frame photon number is assumed to be constant). This coupling is observed in the optimized, shaped configuration (Fig. \ref{FigS3}, bottom), whereas the intensity appears to be essentially decoupled from the mechanical motion in the unshaped case (Fig. \ref{FigS3}, top). 

\section{Photon noise characterization}

\begin{figure}[t!]
\includegraphics[width=85mm]{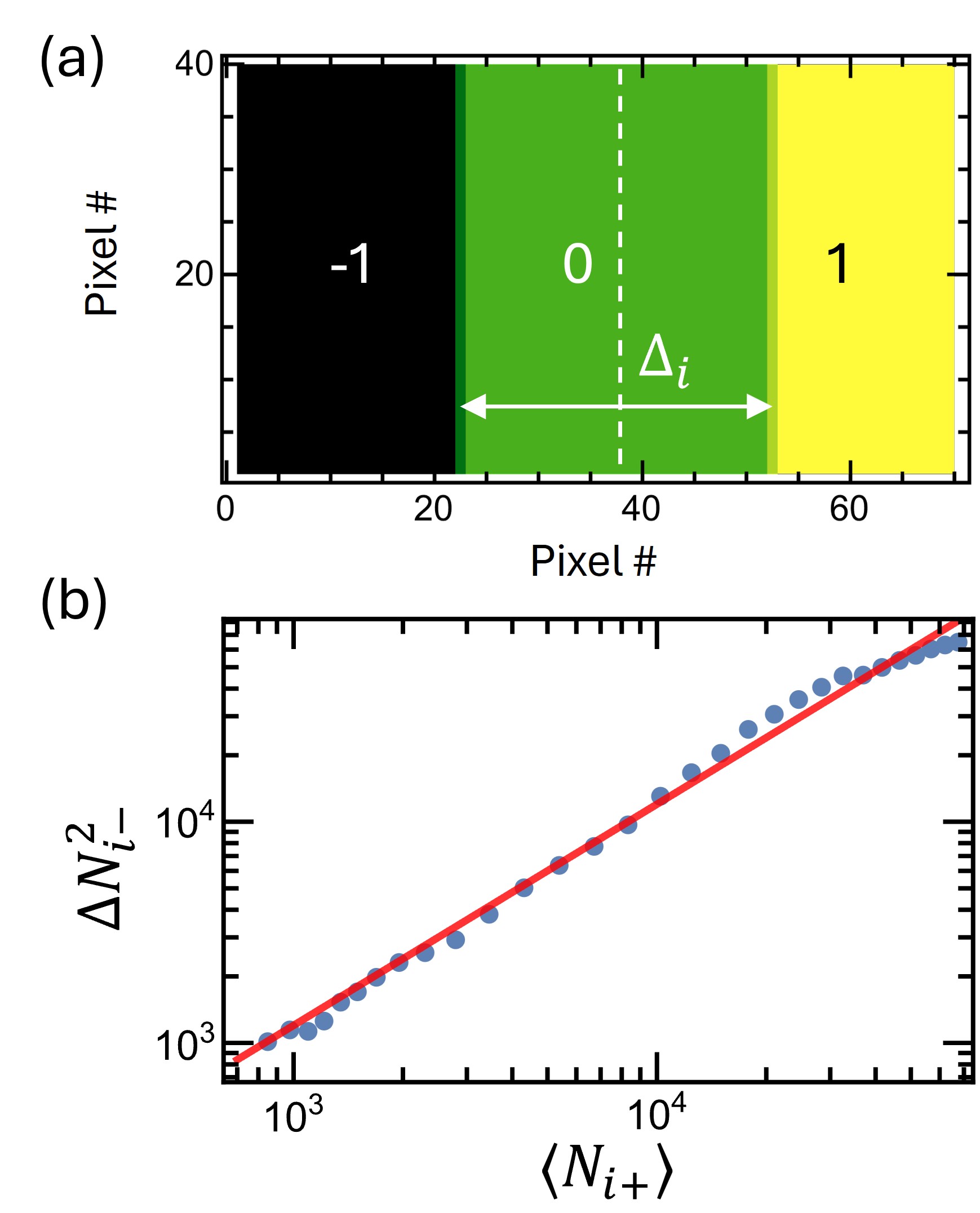} \centering
\caption{\textbf{Photon noise characterization.} The noise variance $\Delta N_{i-}^2$  of a family of differential estimators is compared to their average photon number $\langle N_{i+}\rangle$ (dots), displaying a linear relationship (red, straight line). }%
\label{FigS4}%
\end{figure}

The He-Ne laser used for performing our experiment exhibits sizeable, non-stationary low-frequency amplitude noise. To ensure quantum-limited operation while acquiring the data, we perform a self-consistent characterization of the laser amplitude noise, which essentially consists in constructing a fluctuations vs average photon number diagram, from the same noise measurement dataset as that used to compute the noise estimators of Fig. 5(a) (right column) from the main manuscript. To do so, we build a family of balanced masks $g_{h,i}$ discounting the pixels belonging to a centered window of width $\Delta_i\in\llbracket 1\,\mathrm{px},60\,\mathrm{px}\rrbracket$ (see Fig. \ref{FigS4} (a)). Each of these masks is convoluted with the reference data $N |u_{\mathrm{foc},0}(\vec{\rho},t)|^2$ to obtain the following family of differential estimators:

\begin{eqnarray}
\hat{N}_{i-}(t)&=&N\int\mathrm{d}^2\vec{\rho}\,g_{h,i}(\vec{\rho})|u_{\mathrm{foc},0}(\vec{\rho},t)|^2,\label{eq:S5}
\end{eqnarray}

whose variance $\Delta N_{i-}^2$ is further determined. Likewise, a family of sum operators is obtained following a similar approach:

\begin{eqnarray}
\hat{N}_{i+}(t)&=&N\int\mathrm{d}^2\vec{\rho}\,|g_{h,i}(\vec{\rho})||u_{\mathrm{foc},0}(\vec{\rho},t)|^2,\label{eq:S6}
\end{eqnarray}

whose average $\langle N_{i+}\rangle$ corresponds to the average number of photons contributing to the estimator $\hat{N}_{i-}$. Fig. \ref{FigS4} shows that $\Delta N_{i-}^2$ and the average number of photons $\langle N_{i+}\rangle$ are linearly related, which establishes that the family of differential estimators operates at the shot noise limit. This applies in particular to the wavefront-shaped split estimator noise, and subsequently to the other two estimators displayed in the mid column of Fig. 5(a) from the main manuscript, which show very similar levels of noise.

\bibliography{bib}